\documentclass[a4paper,fleqn,usenatbib]{mnras}

\usepackage{hyperref}
\usepackage{cleveref}
\usepackage{times,txfonts}
\usepackage{graphicx}
\usepackage{pdflscape}
\usepackage[T1]{fontenc}
\usepackage{ae,aecompl}
\pdfoutput=1 

\usepackage{graphicx}	
\usepackage{amssymb}	
\usepackage{bm}

\title[IGR~J17091-3624 over the last decade]{Swift/XRT, Chandra and XMM-Newton observations of IGR J17091-3624 as it returns into quiescence}

\author[Pereyra et al.]{M. Pereyra$^{1}$,  D. Altamirano$^{2}$, J. M. C. Court$^{3}$, N. Degenaar$^{4}$, R. Wijnands$^{4}$, 
\newauthor
 A. S. Parikh$^{4}$ and V. A. C\'uneo$^{5,6}$
\\
$^{1}$ CONACYT, Instituto de Astronom{\'\i}a, Universidad Nacional Aut\'onoma de M\'exico, 22860 Ensenada, BC, Mexico\\
$^{2}$ School of Physics and Astronomy, University of Southampton, B46, Southampton, SO17 1BJ, UK\\
$^{3}$ Department of Physics and Astronomy, Texas Tech University, PO Box 41051, Lubbock, TX 79409, USA\\
$^{4}$ Anton Pannekoek Institute for Astronomy, University of Amsterdam, Science Park 904, NL-1098 XH Amsterdam, the Netherlands \\
$^{5}$ Instituto de Astrof\'isica de Canarias (IAC), E-38205 La Laguna, Tenerife, Spain \\
$^{6}$ Departamento de Astrof\'isica, Universidad de La Laguna, E-38206 La Laguna, Tenerife, Spain \\
}

\date{Accepted XXX. Received YYY; in original form ZZZ}

\pubyear{2020}
\hypersetup{draft}
\begin{document}
\label{firstpage}
\pagerange{\pageref{firstpage}--\pageref{lastpage}}
\maketitle
\begin{abstract} 
IGR J17091-3624 is a low mass X-ray binary (LMXB) which received wide attention from the community thanks to its similarities with the bright black hole system GRS 1915+105. Both systems exhibit a wide range of highly structured X-ray variability during outburst, with time scales from few seconds to tens of minutes, which make them unique in the study of mass accretion in LMXBs. In this work we present a general overview into the long-term evolution of IGR J17091-3624, using {\itshape Swift}/XRT observations from the onset of the 2011--2013 outburst in February 2011 to the end of the last bright outburst in November 2016.  We found 4 re-flares  during the decay of the 2011  outburst, but no similar re-flares appear to be present in the latter one. We studied in detail the period with the lowest flux observed in the last 10 years, just at the tail end of the 2011-2013 outburst, using {\itshape Chandra} and {\itshape XMM-Newton} observations. We observed changes in flux as high as a factor of 10 during this period of relative quiescence, without strong evidence of softening in the spectra. This result suggests that the source has not been observed at its true quiescence so far. By comparing the spectral properties at low luminosities of IGR J17091--3624 and those observed for a well studied population of LMXBs,  we concluded that  IGR J17091--3624  is most likely to host a black hole as a compact companion rather than a neutron star.
\end{abstract}




\section{Introduction}\label{sec:Intro}
Among the whole population of binary systems in the universe, LMXBs are one of the best targets to study the accretion processes occurring in extreme physical regimes. These systems are formed by a low-mass star orbiting a compact object, which can be either a Neutron Star (NS) or a Black Hole (BH), that accretes matter from its low-mass stellar companion by Roche-lobe overflow. Depending on the amount of mass transferred to the compact object, LMXBs will exhibit different accretion states during their lives. The extremes of these accretion regimes are the so-called outburst  and quiescence states, characterized by high and low X-ray luminosity, respectively. During outburst, LMXBs have been observed at very high luminosities in the 0.5 - 10 keV band, within a range of $10^{34}$ - $10^{39}$ erg s$^{-1}$. In the quiescent state, when very low or no mass accretion takes place, their luminosities are as low as  $10^{30}$ - $10^{33}$ erg s$^{-1}$ (e.g. \citealp{Wijnands2015,Wijnands2013,Reynolds2014,Plotkin2013,Heinke2008,Jonker2007,Rem&Mc2006} and references therein). Although many observational properties of LMXBs at high luminosities are relatively well understood, given the current sensitivity of X-ray instruments the spectral behavior of LMXBs at luminosities $\lesssim$ $10^{34}$erg s$^{-1}$ is less clear (\citealp{2013MNRAS.434.1599W,2016RAA....16...62Y,2018RAA....18..142C}).
GRS 1915+105 is one of the well known LMXBs whose behavior at the high luminosity regime is fairly well characterized and extensively studied in outburst (\citealp{Belloni2000}). Since the discovery outburst in 1992 \citep{Castro-Tirado92}, an  exponential decay of its X-ray emission has been observed a few times (\citealp{2018ATel11828....1N}). The source also exhibited an unusual low flux state last year \citep{2019ATel12742....1H}, with short X-ray and bright radio flares (\citealp{2019ATel12848....1B,2019ATel12839....1K,2019ATel13316....1R}, and references therein). More recently, evidence of renewed activity in the microquasar has been observed by {\itshape INTEGRAL} and {\itshape MAXI} instruments \citep{2020ATel13676....1L,2020ATel13652....1A}.
GRS 1915+105 is a very peculiar BH binary that exhibits a complex X-ray variability in time scales from few seconds to tens of minutes. Its X-ray light curve shows several quasi-periodic oscillations at high and low frequencies, often associated with the existence of instabilities in the accretion disk (e.g. \citealp{Janiuk2000}). Over the last decades, only two other LMXBs have exhibited some of the exotic variability observed in GRS 1915+105:  the BH candidate IGR J17091-3624 and the NS MXB 1730-335  \citep{Altamirano2011,Bagnoli2015, 2018A&A...612A..33M}. 
Although both sources have experienced quiescent periods in the past, to characterize their behavior in this state requires detailed observations at very low luminosities which had been only possible for IGR J17091-3624 \citep[see e.g.][]{Wijnands2012}. 

IGR J17091-3624 has shown multiple outbursts from the time of its discovery in 2003 \citep{Rev2003,Capitanio2006}. For the last two of its outbursts, an extensive observing campaign was performed by the current X-ray missions. During these two outbursts at least nine variability classes were identified to resemble those previously seen in GRS 1915+105 \citep{Altamirano2011,Capitanio2012,Pahari2013a,Zhang2014,Court2017}. The presence of extreme winds in the accretion disk, similar to those observed in GRS 1915+105, associated to its quasi-periodic variability, has also been reported  \citep[by][]{King2012,Janiuk2015}. The transient nature of IGR J17091-3624 allowed X-ray missions to also observed it at very low count rates, in the so called quiet periods. The first quiet period was observed by {\itshape XMM-Newton} in 2006 and 2007, with only two observations performed at that time, as reported by \cite{Wijnands2012}. For the second quiet period, at the very tail end of the 2011 outburst, five observations were performed by {\itshape Chandra} and {\itshape XMM-Newton} instruments at different epochs. Additionally, {\itshape Swift}/XRT observations provided a very nice coverage of its transition from outburst to quiescence, making the source particularly interesting to study the low-level accretion in X-ray binaries. 

Although it is well known that a large number of LMXBs spend most of their life in a dormant state, with very weak or zero mass accretion, the definition of this quiescent state in BH binaries is still matter of debate. To define an upper limit for low accretion regimes, parameters like the normalized Eddington X-ray luminosity $l_{x}$ (with $l_{x}$ = $L_{x}$/$L_{Edd}$, where $L_{x}$ is the X-ray luminosity from 0.5 to 10 keV and $L_{Edd}$ = $1.26$ $\times$ $10^{38}$ $[M/M_{\odot}]$ erg s$^{-1}$ for ionized hydrogen) are commonly used \citep{Rem&Mc2006,Plotkin2013}. Under these assumptions, a 10$M_{\odot}$ BH with $l_{x}$$\sim$ $10^{-8.5}$-$10^{-5.5}$, corresponding to $L_{x}$$\sim$ $10^{30.5}$-$10^{33.5}$erg s$^{-1}$, is considered to be in a quiescent state.
The low X-ray flux observed in this regime has been generally associated to radiatively inefficient accretion flows \citep[RIAF; see][ and references therein]{Nara-Yi94,Nara&Mc2008} and the observed correlation between the X-ray flux and the presence of radio jets proposed by \cite{Fender2003,Gallo2006, Miller2008} has also become an interesting scheme to explain the spectral properties of X-ray binaries in quiescence \citep{Miller2011,Plotkin2016}. 
The conditions under which they reach the true quiescent state are not well understood either. Reflaring activity or mini-outburst episodes have been widely detected in many LMXBs, with different timescales and brightness, but very little is know about the physical mechanisms that generate them (see e.g. \citealp{Lasota2001,Patruno2009,Parikh2017b,Patruno2016}). To analyse how the quiescent stage is influenced by the transition from outburst to quiet periods, requires a detailed characterization of the long term variability like the one presented here. We have studied the X-ray emission of IGR J17091-3624, outside the outburst regime,  by following its transition towards quiescence.  We characterised the long-term X-ray variability observed in the source after the main outburst in 2011 and analysed the spectral properties at the lowest luminosity period observed over the last decade. Finally we pointed out the differences observed in the X-ray properties of the source during its last two outbursts and discuss the implications of our results in the context of previous findings for this source and other LMXBs.

\section{Observations} \label{sec:Obs} 
To characterize the time evolution of IGR J17091-3624 X-ray flux during its last two outbursts we used data obtained with the {\itshape Swift}/XRT instrument \citep{2000SPIE.4140...64B}, on board of the Neil Gehrels {\itshape Swift} Observatory \citep{2004ApJ...611.1005G}, performed between February 2011 (MJD 55595) and November 2016 (MJD 57690). To study the transition of the source towards quiescence we selected archived {\itshape Chandra} \citep{1999AAS...195.9601W} and {\itshape XMM-Newton} \citep{1999ESABu.100...15B} observations performed within June - October of 2013. These data, presented here for the first time, correspond to the five last months of the 2011-2013 outburst (for further information about the behaviour of the source in outburst we refer to \citealp{2020MNRAS.491.4857G,2018Ap&SS.363..189D,2017ApJ...851..103X,Court2017,Grin2016} and references therein).

\subsection{{\itshape Swift}} \label{subsec:Swift}
\begin{figure*}
\centering
     \includegraphics[scale=0.30]{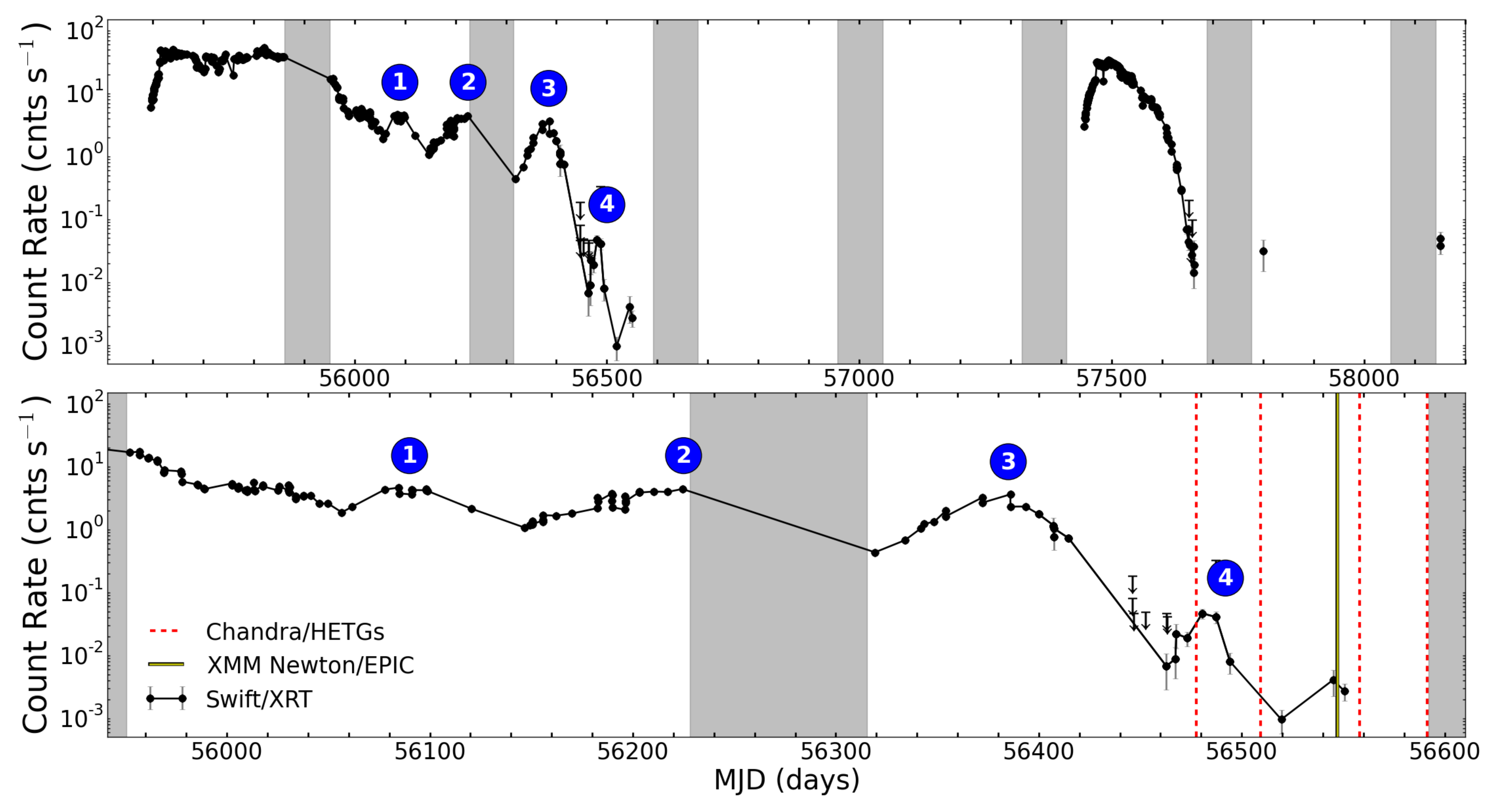}
       \caption{{\itshape Top:} {\itshape Swift}/XRT long-term light curve of IGR J17091-3624 in the 0.3-10 keV band, showing the last two outbursts experienced by the source in February 2011(MJD 55595) and February 2016(MJD 57445). {\itshape Bottom:} Zoom in on the re-flare period of the source observed at the tail end of the 2011 outburst. Four different re-flares are clearly seen, the peaks are marked with blue circles. Black arrows indicate {\itshape Swift/XRT} 3$-\sigma$ upper limits.  Gray shaded areas correspond to the times when the source was not observed by {\itshape Swift}/XRT due to Sun constraints. The dashed red and solid yellow vertical lines in the bottom panel indicate the times when {\itshape Chandra} and {\itshape XMM-Newton} observations were performed, respectively.  Please refer to sections \ref{sec:long_term} and \ref{sec:flares} for a detailed description of this figure.} 
    \label{Swift}
\end{figure*}

{\centering \subsubsection{Light curve}}\label{lightcurve}
We retrieved the X-ray light curve from {\itshape Swift} observations using the {\itshape Swift}-XRT data products generator  \citep[see][]{Evans2009,2007A&A...469..379E}. Considering a snapshot binning method (a single and continuous on-target exposure of 300-2700s) we extracted a 0.3-10 keV long term light curve from MJD 55500 to MJD 58200, which includes the outburst onset in February 2011 (MJD 55595) and the last {\itshape Swift}/XRT observations of IGR J17091-3624 in 2018 (MJD 58151). In the light curve presented in Fig. \ref{Swift} black circles correspond to detections above $3\sigma$. Upper limits are mark with black arrows, usually associated to observations at low count rates with short exposure times ($\leq$ 400s).

{\centering \subsubsection{XRT spectra}} \label{XRT spectra}
The data were downloaded from the HEASARC archive and processed using the {\itshape xrtpipeline} tool. We extracted the spectra using {\itshape XSelect}. A source region of 40 arcsec centred on the source location was used. An annular background region (also centred on the source location) having an inner radius of 150 arcsec and an outer radius of 250 arcsec was used. The ancillary response files were created using the {\itshape xrtmkarf} tool and the appropriate response matrix files, as suggested by the tool, were used. The spectra were binned to have a minimum number of 10 counts per bin for data taken in the Window Timing (WT) mode and 2 counts/bin for data taken in the Photon Counting (PC) mode. 

\subsection{{\itshape XMM-Newton}} \label{subsec:XMM-Newton}
{\centering \subsubsection{EPIC spectra}}
The data were acquired with the European Photon Imaging Camera (EPIC) using the MOS and pn CDD arrays in full frame configuration and the thin optical blocking filter \citep{2001A&A...365L..18S,2001A&A...365L..27T}.
We used a 33.7~ks observation performed on September 12th, 2013 (MJD 56547). 
Observation IDs (OBSIDs), energy range and count rates used for each camera are presented in Table \ref{tab:Observations}.
To obtain the event files lists, we reprocessed the Original Data Files (ODFs) using {\itshape epproc} task from {\itshape XMM-Newton} Science Analysis System (SAS) v.1.2, adopting standard procedures\footnote{https://www.cosmos.esa.int/web/xmm-newton/sas-threads}. Due to high energy particle background events detected at the end of the observation, we excluded the last 10ks from the EPIC/pn data and 4ks from those of the MOS cameras. The source spectrum was extracted with {\itshape xmmselect} using a circular region of 25'' centred on the source. To estimate the background contribution we used a 50'' arcsec annulus region around the source position, with an inner radius of 30'', excluding the source extraction region. Redistribution matrix and ancillary files were created using {\itshape rmfgen} and {\itshape arfgen}, respectively. All the spectra were grouped to contain at least 15 counts per bin. Due to the very low count rates at the time of the observation, data from the Reflecting Grating Spectrometer were not considered for this analysis. 

\begin{table*}
\caption{Log of the {\itshape Chandra} and {\itshape XMM-Newton} observations.}
\label{tab:Observations}
\begin{tabular}{cccccccc}
\hline
{\bf OBS ID} &  {\bf Date}       & {\bf Exp. time} &{\bf  Telescope} & {\bf  Energy Range}  &  {\bf Count Rate}   & {\bf Total}       &     \\
    	    & {\scriptsize (DD/MM/YYY)} &    (ks)  & {\bf Instrument}  		          &    (keV)             &  (10$^{-3}$ cts s$^{-1}$) &  {\bf No. of Counts}  &     \\
\hline
\multicolumn{7}{c}{\bf {\itshape Chandra}} \\
\hline
\hline
&&&&&&& \\
14658   &  04/07/2013  &  39.3	& ACIS/HETGS(HEG)&	0.8 $-$ 10   &  9.1 $\pm$ 0.5  &   	 394	      &	      \\
 	    &    		   &   		& ACIS/HETGS(MEG)&	0.4 $-$ 5    &  11.1 $\pm$ 0.1 &   	 467	      &	      \\
 	    & 	    	   &   		& ACIS/HETGS(Zeroth)&	0.4 $-$ 10   &  29 $\pm$ 1     &   	 1132     &	      \\
&&&&&&& \\ 
14659  &  05/08/2013  &  40.5	& ACIS/HETGS(HEG)&	0.8 $-$ 10   &  0.5 $\pm$ 0.2  &   	 30	      &	       \\
       &  	          & 		& ACIS/HETGS(MEG)&	0.4 $-$ 5    &  1.5 $\pm$ 0.3  &   	 85	      &	       \\
       &	          &   		& ACIS/HETGS(Zeroth)&	0.4 $-$ 10   &  2.9 $\pm$ 0.3  &   	 125  &	      \\
&&&&&&& \\
14660   &   23/09/2013 &  42.2	& ACIS/HETGS(HEG)&	0.8 $-$ 10    &  0.9 $\pm$ 0.2  &   	 56	      &	       \\
  	    & 	    	   &		& ACIS/HETGS(MEG)&	0.4 $-$ 5     &  2.8 $\pm$ 0.3  &   	 138	  &        \\
 	    & 		     &   		& ACIS/HETGS(Zeroth)&	0.4 $-$ 10    &  3.0 $\pm$ 0.3  &   	 138  &	      \\
&&&&&&& \\
14661   &   26/10/2013 &  39.2	& ACIS/HETGS(HEG)&	0.8 $-$ 10    &  2.4 $\pm$ 0.3  &   	 136	      &	       \\
	    &   	       & 		& ACIS/HETGS(MEG)&	0.4 $-$ 5     &  3.1 $\pm$ 0.4  &   	 156	      &	       \\
 	    & 		       &   		& ACIS/HETGS(Zeroth)&	0.4 $-$ 10    &  7.2 $\pm$ 0.4  &   	 296      &	      \\
&&&&&&& \\
\hline
\multicolumn{7}{c}{\bf {\itshape XMM-Newton}} \\
\hline
\hline
&&&&&&& \\
 	       &    	        &  33.7	& EPIC(PN)	 &	0.2 $-$ 15   & 23.9 $\pm$ 0.8  &   	 807	      &	       \\
0721200101 & 	12/09/2013  &  39.8	& EPIC(MOS1) &	0.2 $-$ 15   &  6.1 $\pm$ 0.5  &   	 346	      &	       \\
	       & 		        &  39.2	& EPIC(MOS2) &	0.2 $-$ 15   &  6.9 $\pm$ 0.5  &   	 405	      &	       \\
&&&&&&& \\
\hline
\end{tabular}
\end{table*}

\subsection{{\itshape Chandra}} \label{subsec:Chandra}
\hspace{0.5cm} We retrieved four observations from the {\itshape Chandra} Data Archive. The data were acquired with the Advanced CCD Imaging Spectrometer S-array (ACIS-S; \citealp{2003SPIE.4851...28G}) operating in a Faint Timed Exposure mode, and the High Energy Transmission Gratings (HETGS) in the focal plane with the zeroth-order flux incident in the S3 chip. 
The HETGS on board of {\itshape Chandra} consists of two sets of gratings: the Medium Energy Grating (MEG) and the High Energy Gratings (HEG), optimized for medium energies and high energies respectively.
We created event files using {\itshape chandra\_repro} script from {\itshape Chandra} Interactive Analysis of Observations (CIAO) 4.8 and the Calibration Data Base (CALDB) 4.7.0 software, following standard analysis threads\footnote{http://cxc.harvard.edu/ciao/threads/spectra$\_$hetgacis}. Energy ranges and final count rates for each spectrum are described in Table \ref{tab:Observations}.

{\centering \subsubsection{HETGS zeroth order spectra}}\label{HETGS_zeroth}
\hspace{0.5cm} To extract zeroth-order grating spectra we used \emph{spectextract} CIAO tool, following standar procedures\footnote{http://cxc.harvard.edu/ciao/ahelp/specextract.html}. We used a circular spatial region of 6.5'', centered in the zeroth order location automatically selected by tgdetect in the \textit{chandra$\_$repro} script, to obtain the source spectrum. The ancillary and response files were generated by \emph{specextract} during spectra extraction. All the spectra were grouped to contain at least 15 counts per bin. 

{\centering \subsubsection{HETGS first order spectra}}\label{HETGS_first_order}
\hspace{0.5cm} We extracted pha2 files for both gratings using \emph{tgextract2}\footnote{http://cxc.harvard.edu/ciao/ahelp/tgextract2.html} CIAO tool. The REGION block of the pha2 file was used to extract the source dispersed flux. We estimated the background considering a extraction region of the same size of that used for the source. The background flux was estimated from the events with cross-dispersion contained in the regions lying above and below the source extraction region. The source and background first order spectra from the HEG and MEG, as well as their corresponding redistribution matrix and ancillary files, were obtained using {\itshape tgsplit} and {\itshape mktgresp}, respectively.  To increase the signal to noise of the extracted spectra we combined the -1 and +1 orders for each arm using {\itshape combine\_grating\_spectra}.

\section{DATA ANALYSIS} \label{sec:Analysis} 
\begin{figure*}
     \includegraphics[width=\textwidth]{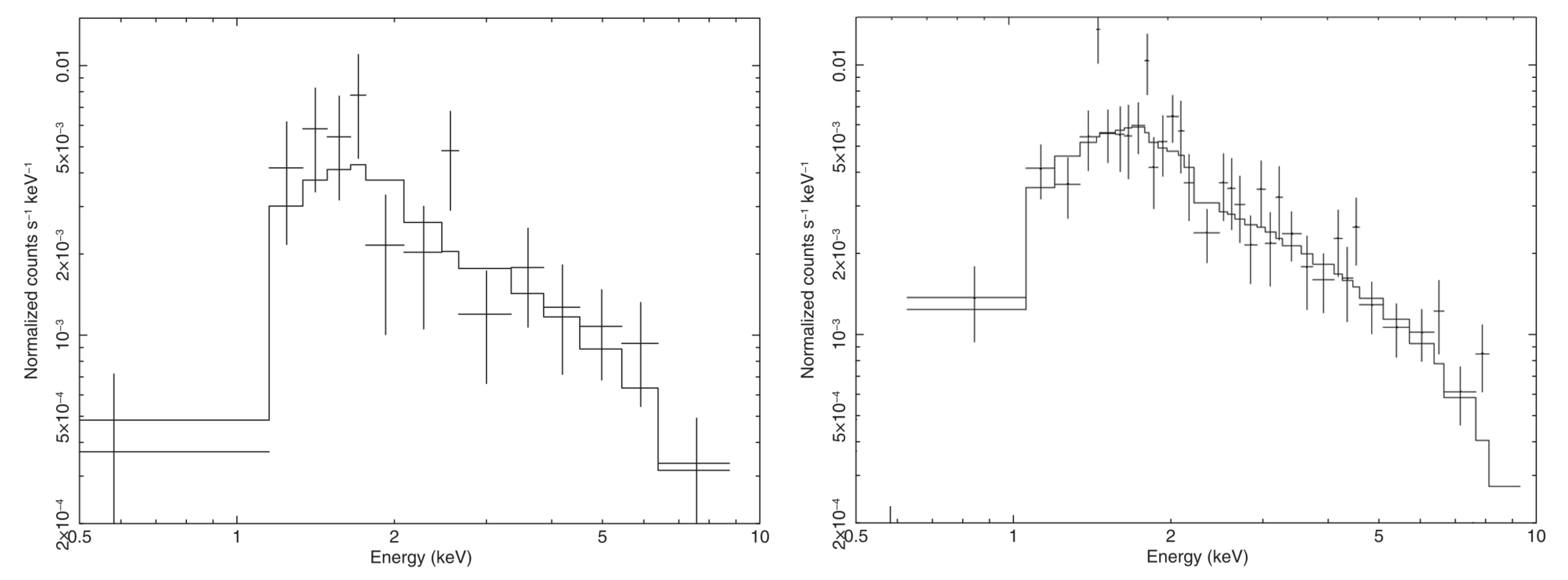}
       \caption{We compare here the {\itshape XMM-Newton} EPIC/pn spectrum from \citet[][left]{Wijnands2012} to the {\itshape XMM-Newton} EPIC/pn spectrum of IGR J17091-3624 at the low luminosity period in 2013 (right), to illustrate the higher statistics obtained from our data set. The solid line through the data points is the best-fitting absorbed power-law model.}
    \label{XMM-Newton_Data}
\end{figure*}

\hspace{0.5cm} The spectral fitting was conducted with {\Large XS}PEC v.12.9.0 \citep{Arnaud1996}. We used $\chi^{2}$ statistics to fit {\itshape Swift}/XRT, EPIC and HETGS zeroth order spectra. When HETGS first order spectra and {\itshape Swift}/XRT spectra with the lowest count rates (2 counts/bin) were considered, we used {\itshape cash} statistics. \\ Only for the EPIC data, the highest quality data, it was possible to analyse five different emission models that are commonly used to describe LMXBs spectra: a power law photon spectrum (POWERLAW), thermal bremsstrahlung (BREMSS), black body radiation (BBODY), accretion disk/multi-black body emission (DISKBB) and a NS hydrogen atmosphere model (NSATMOS). We included the effect of neutral hydrogen absorption N$_{H}$ using the TBABS model Version 1.0 with default {\Large XS}PEC abundances (\citealp{2000ApJ...542..914W}) and cross-sections (\citealp{1996ApJ...465..487V}). To give the reader an idea of the quality of our data at the lowest count rates, the best-fitting for EPIC/pn spectra is presented in Fig. \ref{XMM-Newton_Data} compared to previous studies data at similar count rates \citep{Wijnands2012}. \\
The X-ray flux was calculated considering different energy ranges, according to the instruments responses, using the convolution model CFLUX. Since the distance to the source is unknown, to estimate its luminosity we considered distances commonly used in previous works, which place the source between 8 kpc and 35 kpc. The errors reported here are at a 90$\%$ confidence level unless otherwise stated. In the following sections, we present a detailed description of the spectral analysis performed for each data set individually.

{\centering \subsection{{\itshape Swift}/XRT data}} \label{subsec:Swift/XRT_fits}
From the light curve presented in Fig. \ref{Swift} we see that there is a remarkable difference in the flux level of {\itshape Swift}/XRT data along the transition period, from outburst to quiescence, that have to be considered carefully. Although we confirmed that for the highest count rates was feasible to use two component models (black body + power law) to fit the data, this was not necessarily the case for those data sets with the lowest count rates. By using the same models to fit both types of data sets (high and low count rates), we found that for the low count rates data the model fitted always becomes insensitive to the additional model component. According to {\Large XS}PEC manual this result means that the fit is unable to constrain the parameter and it should be considered indeterminate, which usually indicates that the model is not appropriate. In order to be consistent in comparing the spectral properties of the X-ray emission at different flux levels, we decided to use only single component models to fit all {\itshape Swift}/XRT observations along the decay period of IGR J17091-3624. Under this assumption, we found that an absorbed power-law spectra fits consistently to all {\itshape Swift}/XRT observations. From this model we estimate fluxes in the 0.7--10 keV energy range for the WT mode data\footnote{
http://www.swift.ac.uk/analysis/xrt/digest$\_$cal.php} and the 0.5--10 keV energy range for the PC mode data. Both the $N_\mathrm{H}$ and the photon index ($\Gamma$) were free to vary during the fitting as this was necessary to compare with the results reported by \cite{Wijnands2015}. Data for which the error bars on $\Gamma$ exceeded 0.5 were discarded. The results are discussed in sections \ref{sec:flares} and \ref{sec:Dis_Spectral_evolution}. \\

{\centering \subsection{{\itshape XMM-Newton}/EPIC data}\label{XMM-Newton_fits}}
With the {\itshape XMM-Newton} observations, corresponding to the lowest flux observed from the source during the 2011-2013 outburst decay, we were able to investigate if the source actually reached quiescence by considering the most commonly used models for LMXBs at this stage: BREMSS and POWERLAW spectral shapes. We fitted the three EPIC spectra simultaneously, for the 0.5-10 keV energy range, with N$_{H}$ value linked between observations and all the parameters varying freely during the fitting. We used independent normalization parameters for the {\itshape MOS} and {\itshape pn} CCD cameras. For the BREMSS spectral shape no good fit was found, the model errors were always above 50\% of the estimated value. We obtained the best-fitting to our data using the absorbed {\itshape POWERLAW} model with N$_{H}$ = 1.1 $\pm$ 0.2 $\times$ $10^{22}$ cm$^{-2}$, in excellent agreement with the outburst value reported by \cite{Rodriguez2011} from  {\itshape Swift}/XRT and {\itshape INTEGRAL}/ISGRI data. We estimated an absorbed flux value of  1.4 $\pm$ 0.2 $\times$ 10$^{-13}$ erg s$^{-1}$ cm$^{-2}$ in the 0.5 - 10 keV energy band, with a hard photon index $\Gamma$ = 1.6 $\pm$ ${0.1}$, consistent with previous findings for the quiet period in 2006-2007 from \cite{Wijnands2012}. \\
We also used EPIC data to investigate the possible contribution of a thermal component to the X-ray spectra, expected to be observed in neutron star LMXBs in quiescence but not for the case of BH systems (\citealp{2001nsbh.conf..245B}). From all observations at the lowest count rates, only EPIC/pn data were considered for this attempt given its higher count rates and statistical significance compared to EPIC/MOS and {\itshape Chandra}/HETGS data sets. For EPIC/pn spectra we added a thermal component to the POWERLAW fitting, considering BBODY, DISKBB and NSATMOS spectral shapes, independently. Estimated errors were above 50\% of the estimated value for the temperature of the thermal component in all cases, likely caused by the fit being insensitive to the parameter. An F-test applied to compare POWERLAW with multiple component models yields a probability value of  0.85, 0.98 and 0.74 for the BBODY+POWERLAW, DISKBB+POWERLAW and NSATMOS+POWERLAW models, respectively, which indicates that adding an extra model to the fitting does not improve the modeling of the spectra. Similar to our {\itshape Swift}/XRT results, for data with the lowest count rates, we conclude that multiple component models are not appropriate to fit these observations. 

{\centering \subsection{{\itshape Chandra}/HETGS data}} \label{subsec:Chandra_fits}
\hspace{0.05cm} From each {\itshape Chandra} observation we obtained three different spectra: 1 zeroth order spectrum and 2 more spectra from each arm of the HETGS (first order of HEG and MEG). To account for the limited quality of the first order spectra, we performed a two-step spectral fitting as follows:
\begin{itemize}
    \item[\textit{i.-}] We first fit the 3 HETGS zeroth order spectra simultaneously, considering a {\itshape powerlaw} model with N$_{H}$ value linked between observations and all the parameters varying freely during the fitting. We obtained an N$_{H}$ = 1.1 $\pm$ 0.2 $\times$ $10^{22}$ cm$^{-2}$, consistent with that obtained from {\itshape XMM-Newton} observations (our higher quality data).
    \item[\textit{ii.-}] To better constrain the photon index values in the models, we add the first order spectra to the fitting. Given the low count rates of the first order spectra, we used {\itshape cash} statistics to fit the data in this case, fixing N$_{H}$ to 1.1 $\times$ $10^{22}$ cm$^{-2}$ and fitting simultaneously the 3 spectra (1 zeroth order and 2 first order spectra). 
    \item[\textit{iii.-}] We performed the model fitting for each observation, independently. 

\end{itemize}
 
The log of the observations for all {\itshape Chandra}/HETGS and {\itshape XMM-Newton} data sets, as well as the results for the best-fitting model ({\itshape powerlaw}), are presented in Table \ref{tab:Observations} and \ref{tab:Preliminar_Results: Phabs*Power-Law Model}.

\begin{table*}
\caption{Model fitting results for the low luminosity period, at 90\% level of confidence.}
\label{tab:Preliminar_Results: Phabs*Power-Law Model}
\begin{tabular}{ccccccc}
\hline
{\bf MJD}  	& {\bf  OBS ID} & {\bf No. Spectra} & {\bf N$_{H}$} 		& {\bf  {\bf $\Gamma$} }  &     {\bf F$_{abs}$}$^\dagger$   		  & {\bf  F$_{unabs}$}$^\dagger$                  \\
    		&	       &  {\bf   Fitted}    & (10$^{22}$ cm$^{-2}$)   &	            & (10$^{-13}$erg s$^{-1}$ cm$^{-2}$) & (10$^{-13}$erg s$^{-1}$ cm$^{-2}$)    \\
\hline
\multicolumn{7}{c}{\bf {\itshape Chandra}} \\
\hline
\hline
		56478 & 14658 &        3      &  Fixed to 1.1  &  1.2 $\pm$ 0.1   &	        25 $\pm$ 1		   &	        31 $\pm$ 1		    \\
&&&&&& \\
		56509 & 14659 &        3      &  Fixed to 1.1  &  1.6 $\pm$ 0.3   &	     2.4 $_{-0.7}^{+0.6}$	   &		3.5 $_{-0.4}^{+0.5}$	    \\
&&&&&& \\
		56558 & 14660 &        3      &  Fixed to 1.1  &  1.9 $\pm$ 0.2   &	     2.5 $_{-0.3}^{+0.4}$	   &	        4.2 $\pm$ 0.5	    \\
&&&&&& \\
		56591 & 14661 &        3      &  Fixed to 1.1  &  1.3 $\pm$ 0.2   &	     6.1 $_{-0.6}^{+0.7}$   	   &  		8.0 $_{-0.6}^{+0.7}$	    \\

\hline
\multicolumn{7}{c}{\bf {\itshape XMM-Newton}} \\
\hline
\hline
      56547 & 0721200101&   3       & 1.1   &  1.6 $\pm$ 0.1   &	        1.4 $\pm$ 0.2		   &		2.3 $_{-0.6}^{+0.3}$       \\
\hline
\multicolumn{7}{l}{\footnotesize{$^\dagger$ Flux values correspond to the 0.5-10 keV energy band.}}
\end{tabular}
\end{table*}

\section{Results} \label{sec:Results}

\hspace{0.5cm} We present the {\itshape Swift}/XRT long term X-ray light curve of IGR J17091-3624 from February 2011 to February 2018 in the {\itshape top panel} of Fig. \ref{Swift}. The light curve clearly shows the last two outbursts of the source and the low luminosity period following the 2011. We also identify 4 re-brightening events or {\itshape re-flares} in the light curve. A zoom in on the light curve is presented at the {\itshape bottom panel} of Fig. \ref{Swift}, showing the transition of IGR 17091-3624 from the 2011 outburst towards the lowest luminosity period of the source in 2013. The green and red lines indicate the times at which {\itshape XMM-Newton} and {\itshape Chandra} observations were performed, respectively. We used these five observations, one from {\itshape XMM-Newton} and four from {\itshape Chandra}, to characterize the X-ray emission of IGR 17091-3624 at the lowest count rates observed in 2013. \\

\subsection{Long term evolution} \label{sec:long_term}
The first {\itshape Swift}/XRT detection of IGR J17091-3624 in 2011 corresponds to the outburst onset on 3th of February (MJD 55595). After 22 days (MJD 55616) the flux increased to its maximum value and remained approximately constant for the next 245 days, with an average count rate of $\sim$ 38 cnts s$^{-1}$.  Due to Sun constraints, from October 26, 2011 (MJD 55860) no observations were performed by {\itshape Swift} until after January 26, 2012 (MJD 55952). At this point we observed a steady decrease in the X-ray emission, which reached a minimum count rate of $\sim$ 1.8 cnts s$^{-1}$ on May 10, 2012 (MJD 56057). We identified 4 re-flares peaking at June 2012, October 2012, April 2013 and July 2013, corresponding to MJDs 56098, 56225, 56385, and 56478, respectively. Before the 2016 outburst onset, from the last {\itshape Swift}/XRT detection of IGR J17091-3624, we estimated the lowest count rate for this source to be $\sim$ 10$^{-2}$ cnts s$^{-1}$ in September 10, 2013 (MJD 56545). \\

In February 26, 2016 (MJD 57444) evidence of renewed activity in IGR J17091-3624 was detected by {\itshape Swift}/BAT \citep{MillerATel}, which implies a quiet period of $\sim$ 900 days. The flux of the source increased gradually, in the following 26 days, reaching a maximum count rate of $\sim$ 34 cnts s$^{-1}$ on March 23 (MJD 57470). After May 09, 2016 (MJD 57517) the outburst decay period started with a fast decrease in the source flux observed from May to October 2016 (MJD 57517 to MJD 57663), just before the next Sun constrained period. The last {\itshape Swift}/XRT observation of IGR J17091-3624 was performed on February 3rd, 2018 (MJD 58152) with $\sim$ 3.5 $\times$ 10$^{-2}$ cnts s$^{-1}$, being the lowest count rate reported for its 2016 outburst. \\

We found that during the 2011-2013 outburst the X-ray emission of IGR J17091-3624 remained at a relatively high flux level, above outburst onset value ($\sim$5 $\times$ 10$^{-10}$ erg cm$^{-2}$ s$^{-1}$), for approximately 7.5 times longer than it did in 2016. We estimated similar peak count rates for both outbursts at their highest luminosity period ($\sim$38 cnts s$^{-1}$).  
However, the flux's decay was steeper in the 2016 outburst. We note that the flare-like behavior is only observed during the decay of the 2011-2013 outburst, but we are aware that no continuous observations were performed by {\itshape Swift}/XRT between October 4th, 2016 (MJD 57665) and February 3rd, 2018 (MJD 58152), when the source might have exhibited this variability again. Moreover, the count rate of $\sim$ 1.8 cnts s$^{-1}$ at the beginning of the re-flaring period in the 2011-2013 outburst is 2 orders of magnitude higher than the lowest count rate observed following the decay period of the 2016 outburst ($\sim$ 2 $\times$ 10$^{-2}$ cnts s$^{-1}$). If any re-flare occurred after the observed decay period of the 2016 outburst, they would have started at significantly lower flux levels  ($\sim4\times$ 10$^{-12}$ erg cm$^{-2}$ s$^{-1}$ from the last outburst, compared to the $\sim7\times$ 10$^{-10}$ erg cm$^{-2}$ s$^{-1}$ of the former). We summarize these results in Table \ref{tab:re-flares}.

\begin{table*}
\caption{X-Ray fluxes (0.5-10 keV) estimated for the re-flares. The results are reported at 90\% level of confidence. General properties for the two main outburst experienced by the source, before and after the re-flare period, are also listed to add context to the long term evolution.}
\label{tab:re-flares}
\begin{tabular}{cccccc}
\hline
{\bf MJD}	 & {\bf Telescope/} & {\bf Notes} 	&  {\bf  F$_{unabs}$}                 & {\bf $\Gamma$} & {\bf $\boldsymbol{\chi^2}$/d.o.f.} \\
                &  {\bf Instrument}  &	    	& (10$^{-10}$erg s$^{-1}$ cm$^{-2}$)  &  {\bf }   &  \\ \hline
\hline
\multicolumn{6}{c}{\bf MAIN OUTBURST 2011} \\
\hline
\hline
55595 &  {\itshape Swift}/XRT	&	3rd February 2011, outburst onset             & 2.55 $\pm$ 0.20  &  1.7 $\pm$ 0.05 & 771/795 \\
55616 &  {\itshape Swift}/XRT	&	24th February 2011, outburst maximum flux     & 22.1 $\pm$ 0.20  &  2.5 $\pm$ 0.02 & 1047/814 \\
55952 &  {\itshape Swift}/XRT	&	January 2012, after sun constraints period     & 6.92  $\pm$ 0.1 &  2.4 $\pm$ 0.02 & 925/691 \\
56057 &  {\itshape Swift}/XRT	&	May 2012, minimum flux after main outburst    & 1.44  $\pm$ 0.1 &  2.0 $\pm$ 0.10 & 428/528 \\
\hline
\hline
\multicolumn{6}{c}{\bf RE-FLARING PERIOD 2012-2013} \\
\hline
\hline
56098 &  {\itshape Swift}/XRT	&	June 2012, 1st re-flare (peak)                  & 3.50  $\pm$ 0.10 &  1.7  $\pm$ 0.04 & 446/399	\\
56140 &  {\itshape Swift}/XRT	&	August 2012, 1st re-flare (minimum)             & 1.06  $\pm$ 0.02 &  1.5  $\pm$ 0.10 & 254/212	\\
56225 &  {\itshape Swift}/XRT	&	October 2012, 2nd re-flare (peak)               & 3.99  $\pm$ 0.25 &  1.6  $\pm$ 0.10 &	72/55 \\
56320 &  {\itshape Swift}/XRT	&	January 2013, 2nd re-flare (minimum)            & 0.40  $\pm$ 0.04 &  1.3  $\pm$ 0.20 & 22/19 \\
56385 &  {\itshape Swift}/XRT	&	April 2013, 3th re-flare (peak)                 & 1.69  $\pm$ 0.10 &  1.5  $\pm$ 0.10 & 51/39 \\
56414 &  {\itshape Swift}/XRT	& 	May 2013, 3th re-flare (minimum)                & 0.66  $\pm$ 0.07 &  1.2  $\pm$ 0.20 & 30/32 \\
56467 &  {\itshape Swift}/XRT	& 	Upper limit from {\itshape Swift}/XRT detections in 2013$^\dagger$  & $\leq$ 0.01      &  ---             &  \\
56478 & {\itshape Chandra}/HETGS	&	July 2013, 4th re-flare (peak)          &       0.031 $\pm$ 0.001        &	1.2	$\pm$ 0.1 & 54/65 \\
56547 & {\itshape XMM-Newton}/EPIC &	September 2013, 4th re-flare (minimum)  &  0.0023 $_{-0.0006}^{+0.0003}$ &	1.6 $\pm$ 0.1 & 46/46\\
\hline
\hline
\multicolumn{6}{c}{\bf MAIN OUTBURST 2016} \\
\hline
\hline
57444 &  {\itshape Swift}/XRT	&	February 2016, outburst onset                    &  2.25 $\pm$ 0.02    &   1.6 $\pm$ 0.03 & 785/796 \\
57470 &  {\itshape Swift}/XRT	&	March 2016, outburst maximum flux                &  13.9 $\pm$ 0.10    &  2.2  $\pm$ 0.01 & 1289/872 \\
57610 &  {\itshape Swift}/XRT	&	August 2016, minimum flux after main outburst    &  0.69 $\pm$ 0.02  &  1.8 $\pm$ 0.1 & 682/695  \\
57640 &  {\itshape Swift}/XRT	& 	Upper limit from {\itshape Swift}/XRT detections in 2016$^\dagger$    & $\leq$ 0.01       &  ---         &           \\
\hline
\multicolumn{6}{l}{\footnotesize{$^\dagger$ Upper limits correspond to {\itshape Swift}/XRT spectral fitting that yields errors in $\Gamma$ above 30$\%$ of the estimated value.}} \\
\end{tabular}
\end{table*}

\subsection{Re-flaring activity} \label{sec:flares}
\begin{figure*}
	\includegraphics[clip,scale=0.29]{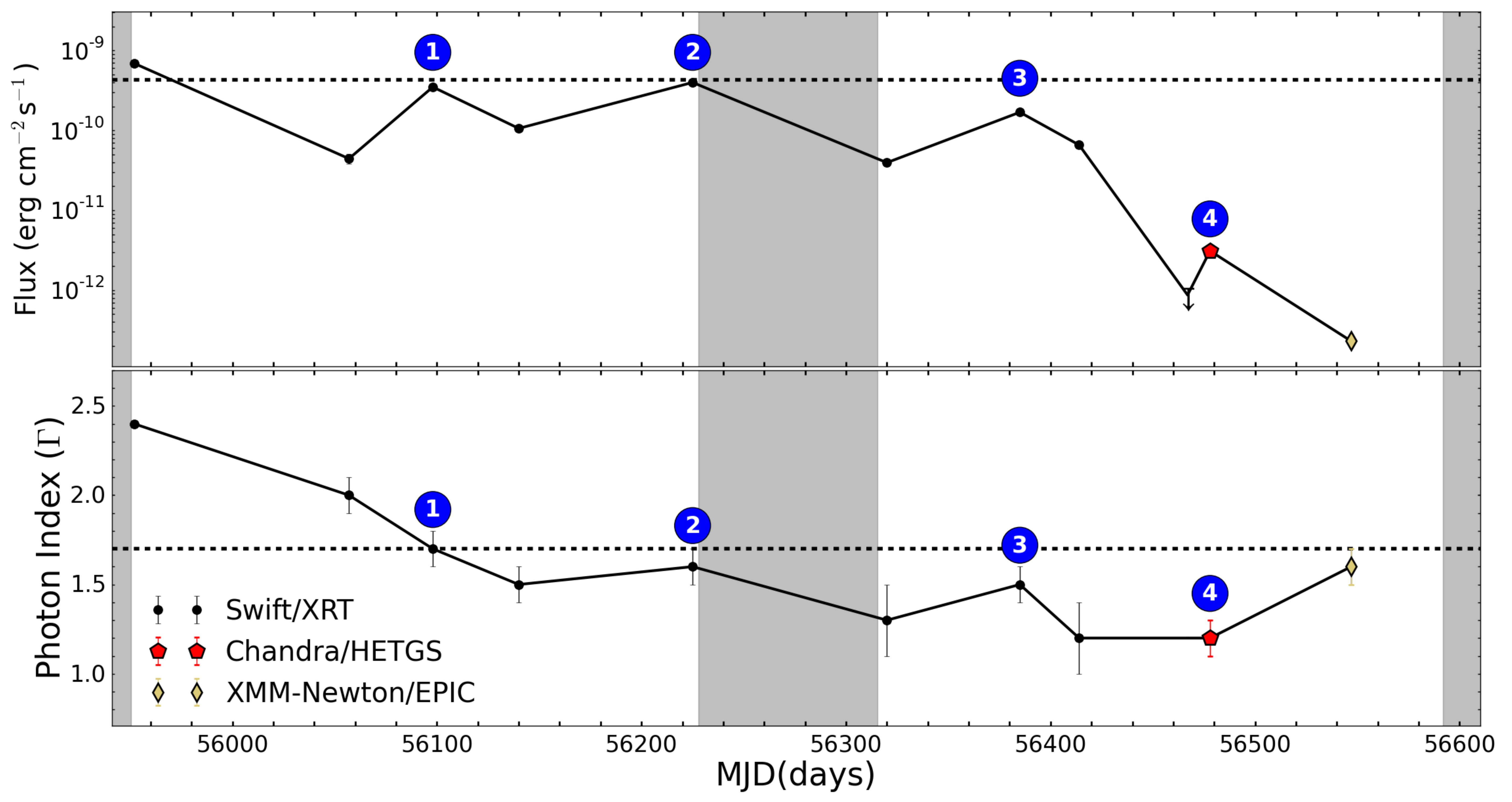}
    \caption{{\bf Tail end of the 2011-2013 outburst.} We present the estimated fluxes  ({\itshape top panel}) and photon index values ({\itshape bottom panel}) for IGR J17091-3624 at the last part of its 2011-2013 outburst. {\itshape Swift}/XRT, {\itshape Chandra}/HETGS and {\itshape XMM-Newton}/EPIC data are marked with black dots, red pentagons and yellow diamonds, respectively. Sun constraints periods are indicated by shaded gray areas. The black arrow marks the upper limit for the flux obtained from {\itshape Swift}/XRT spectra with the lowest count rates. 
    The dashed line indicates the flux level (4.3 $\times$ 10$^{-10}$ erg cm$^{-2}$ s$^{-1}$) and photon index value ($\Gamma$ = 1.7) reported by \citealp{Krimm2011} for the outburst onset in 2011. The flux of the source was found to be always below the outburst onset value along the whole re-flaring period, with no significant evidence of temporal evolution in the spectral shape.}
    \label{re-flares}
\end{figure*}

\hspace{0.5cm}  For the study of the re-flares properties, whose detailed analysis is out of the scope of this work, we only focus on the periods when the higher and lower count rates were observed for each re-flare independently. We assumed that these periods represent the maximum and minimum change in the photon index $\Gamma$, or spectral shape, for the corresponding change observed in the X ray flux of the source as it fades to relative quiescence. In Fig. \ref{re-flares} we present the re-flaring period experienced by IGR J17091-3624 at the tail end of its 2011-2013 outburst, between May 2012-October 2013 (MJD 55952 to MJD 56595), using the X-ray fluxes ({\itshape top panel}) and photon index values ({\itshape bottom panel}) obtained from model fitting described in Section \ref{sec:Analysis}. We plot and analyse the four re-flares, identified in Fig. \ref{Swift}, by taking the highest and lowest values from the fluxes estimated for each re-flare independently. The data used for this purpose are presented in Table \ref{tab:re-flares}. We found the time difference between the peaks of the re-flares ($\Delta$t) to be $\sim$ 127, $\sim$ 160 and $\sim$ 93 days. We found no evident periodicity in the occurrence of these re-flares, but we note that the time period between the 2nd and 3rd re-flares was not properly covered due to a Sun constraint and that 4 events might be too small a sample to look for a periodicity. We note that during the reflaring activity the flux level of IGR J17091-3624 remains always below the value reported by \cite{Krimm2011} at the outburst onset in 2011 (indicated by the black dotted line). For the X-ray emission of the source in the  0.5-10 keV band, we found that fluxes on the peaks of the first three re-flares are 2 orders of magnitude higher than the X-ray flux obtained from the peak of the 4th re-flare.
Although changes in the flux level are evident in the light curve, changes in the spectral shape of the X-ray emission are not that clear. We estimated an average photon index value ($\overline \Gamma$) for the highest and lowest flux values reached during the 4 re-flares. We found $\overline \Gamma$ = 1.5 $\pm$ 0.1 for the peaks, while $\overline \Gamma$ = 1.4 $\pm$ 0.2 for the lowest fluxes. We conclude that there are not significant changes in the photon index along the re-flare period, despite the observed changes in flux seen on each re-flare. 

\subsection{The low luminosity period}
\hspace{0.5cm} After the 3rd re-flare, the X-ray emission of the source dropped significantly and {\itshape Swift}/XRT spectral modelling  only yields an upper limit for the flux of the source. Once the X-ray flux increased during the 4th re-flare, it was possible to model the spectra with a POWERLAW model. However, the errors on the values obtained were large (above 30\% of the estimated value, with errors in $\Gamma$ $\geq$ 0.5) and therefore we decided to excluded these data (all observations in the period MJDs 56420-56570) from our spectral analysis.

\begin{figure*}
	\includegraphics[clip,scale=0.29]{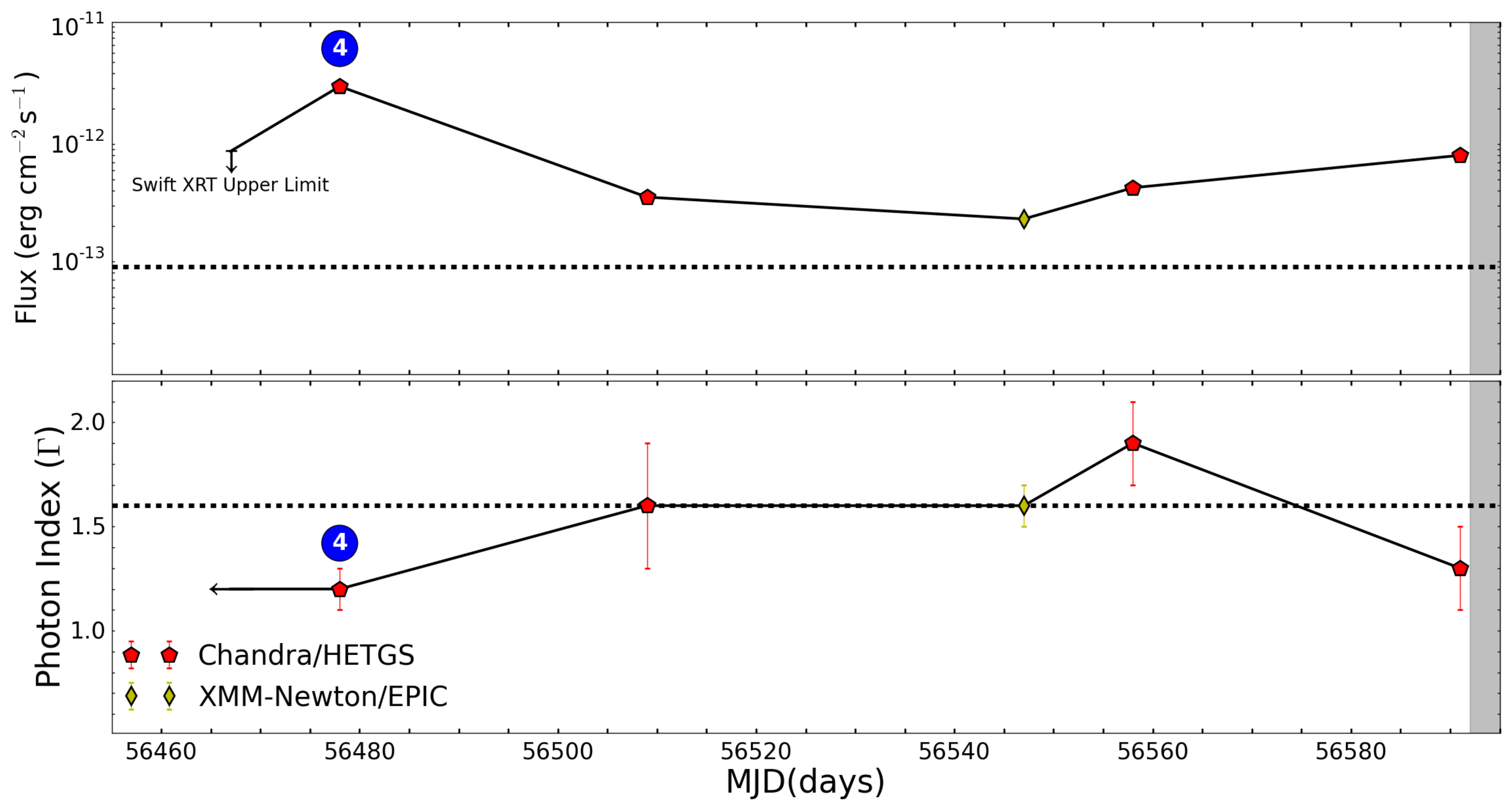}
    \caption{{\bf The low luminosity period.} We present the estimated fluxes  ({\itshape top panel}) and photon index values ({\itshape bottom panel}) for IGR J17091-3624 at the very end of its 2011-2013 outburst. {\itshape Swift}/XRT, {\itshape Chandra}/HETGS and {\itshape XMM-Newton}/EPIC data are marked with black dots, red pentagons and yellow diamonds, respectively. Sun constraints periods are indicated by shaded gray areas. Black arrows mark the upper limits from {\itshape Swift}/XRT detections. 
    The dashed lines indicate the flux level (9$\pm$5 $\times$ 10$^{-14}$ erg cm$^{-2}$ s$^{-1}$) and photon index value ($\Gamma$ = 1.6) reported by \citet{Wijnands2012} for the previous quiescent period of the source in 2006. Slightly variability is observed in the X-ray flux but we did not find strong evidence of softening in the spectra.}
    \label{Figure_Low_Lx}
\end{figure*}

We instead analysed 4 {\itshape Chandra}/HETGS and 1 {\itshape XMM-Newton}/EPIC observations that were performed with higher sensitivity in the same period of time. The data cover a 5 month period, between June and October 2013 (MJD 56477 to MJD 56592), 2.2 years after the 2011-2013 outburst onset in February 2011 and 2.3 years before the source renewed outburst activity in February 2016. 
We present in Fig. \ref{Figure_Low_Lx} the temporal evolution of the X-ray emission of the source during this period. {\itshape Chandra} and {\itshape XMM-Newton} observations are labelled in red and green, respectively. In order to put our results in the context of previous findings, we indicate with a purple dashed line the lowest flux level and photon index value reported by \cite{Wijnands2012} for the previous low luminosity period of the source observed in 2006 and 2007 (9$\pm$5 $\times$ 10$^{-14}$ erg s$^{-1}$ cm$^{-2}$, $\Gamma$ = 1.6 $\pm$ 0.5).
In the {\itshape top panel} we present our flux estimates, showing the peak of the 4th re-flare experienced by IGR J17091-3624 at MJD 56478 (July 5th, 2013), and the subsequent decay of its X-ray emission. From the {\itshape Chandra} observation performed on MJD 56478, at the peak of the 4th re-flare, we estimated a flux value of $\sim$ 3.1 $\times$ 10$^{-12}$ erg s$^{-1}$ cm$^{-2}$. In the following 70 days the flux decreases by a factor of 10, when the source reached its lowest observed flux on September 2013, with a value of $\sim$ 0.23 $\times$ 10$^{-12}$ erg s$^{-1}$ cm$^{-2}$ in the 0.5 - 10 keV band. In the following 40 days the flux of the source increased again, this time only by a factor of 5, as seen by the last {\itshape Chandra} observations before the Sun constrained period started in October 2013.  Since no observations were available after MJD 56600, we cannot discard the possibility that the observed increase in the flux is related to a new re-flare happening at the very end of the 2011-2013 outburst.

The corresponding photon index values obtained from these five observations are shown in the {\itshape bottom panel} of Fig. \ref{Figure_Low_Lx}. Although we observed a factor of 5 to 10 change in the flux levels, corresponding changes in the photon index are not evident. We investigated the statistical significance of this result by performing an F-test to the spectral fitting. To guaranty a statistically meaningful F-test we only considered the data sets with highest quality: {\itshape XMM-Newton}/EPIC/pn and {\itshape Chandra}/HETGS/zeroth-order. We fitted the five spectra simultaneously, using a single component \emph{power-law} model with the photon index varying freely, fixed N$_{H}$ = 1.1 and free normalization parameters to allow differences in the flux level. If the photon index is set to vary freely, but tied between observations, we obtain a constant photon index value of $\sim$ 1.4$\pm$0.1 with $\chi$$^{2}$$/$dof = 145.53$/$144. When we set the photon index to vary independently per observation, we obtain different photon index values (see Table \ref{tab:Preliminar_Results: Phabs*Power-Law Model}) and a $\chi$$^{2}$$/$dof of 121.28$/$140. We then apply an F$-$test on these results and obtain a probability value of $3.9\times10^{-5}$, suggesting that the spectral shape does not remain constant as the source fades into quiescence. However, we found no  evidence of a systematic softening of the spectra. The $\Gamma$ = 1.6 $\pm$ 0.1 obtained for the lowest flux observed in the low luminosity period is consistent to the mean value estimated for the re-flares ($\overline \Gamma$ = 1.45 $\pm$ 0.15) at higher luminosities. It is interesting to note that spectral softening in the spectra of LMXBs is expected at low luminosities, regardless of the nature of its compact object.  For the black hole systems the softening seems to level off at a photon index $\sim$ 2, while the neutron stars reach photon indices of 2.5 - 3 \citep{Plotkin2013,Wijnands2015}. The relatively hard spectra we found in the case of IGR J17091-3624, even at its lowest luminosities, might favor the idea that the source did not reach true quiescence in 2013.

\section{Discussion} \label{sec:Discussion}

\subsection{Outbursts properties}

\hspace{0.5cm} During the last two outbursts the source's X-ray flux reached similar maximum values but we found two main differences between them:\\ 

{\itshape i).- Time scales:} The high flux levels in the 2011-2013 outburst lasted longer than those in 2016. The decay of the X-ray emission to the lowest count rates at the tail end of each outburst was slower for the former. \\

{\itshape ii).- Re-flaring:} Re-flares are clearly identified in the long-term light curve for the 2011-2013 outburst. Three re-flares occurred during the decay of the source's emission towards lower count rates and a 4th re-flare was observed even at its lowest luminosities, while no such re-flaring was detected in the 2016 outburst. \\

We note that there are no observations from {\itshape Swift}/XRT after November 2016 and we cannot rule out that the source shows re-flaring afterwards. Given that the flux level at the end of the 2016 outburst is $\sim$ 2 orders of magnitude lower than the flux observed at the beginning of the re-flare period in 2011-2013, we conclude that any subsequent re-flaring activity in 2016 would be weaker than the one observed in the former outburst. Moreover, it has been shown by some authors that many X-ray transients show different outburst profiles and that re-flare events are not necessarily active after each outburst in LMXBs (\citealp{Parikh2017a,Capitanio2015}). \\ In this context, the absence of re-flares after the 2016 outburst would not be an unusual behavior for a LMXB like IGR J17091-3624 either. \\

\subsection{\textbf{The re-flaring activity}} \label{sec:Dis_Reflaring}
What makes the re-flaring activity in LMXBs still puzzling nowadays is that recent studies show that several sources do actually transit from hard to soft states and back during the re-flares (e.g. MAXI J1535-571 and GRS 1739-278; \citealp{Yan2017,Parikh2019,Cuneo2020}). 
For these sources it's been argued that the transition hard-to-soft/soft-to-hard during the re-flare would depend on the transition luminosity of the main outburst. Only when the luminosity at the re-flare reaches the level at which the hard-to-soft transition happened for the main outburst, the source will transit to the soft state during the re-flare, otherwise it will not. In Fig. \ref{re-flares} we showed that the flux obtained for the re-flares of IGR J17091-3624 never went above the flux level reported for the outburst onset, which is actually below the flux level of the hard-to-soft transition \citep{Krimm2011,2011ATel.3179....1S}. This might explain why the source did not transit back to the soft state, showing no change in the photon index values, in any of the re-flares. We are aware that simultaneous and quasi-simultaneous X-Ray and Radio/NIR/Optical/UV follow up of IGR J17091-3624 re-flares are needed to better constraint their properties, but this analysis is out of the scope of the present work.

\subsection{Black hole or neutron star system}  \label{sec:Diss_BH_vs_NS}

\hspace{0.5cm} To accurately distinguish between a BH and a NS LMXB often requires direct measurements of the physical properties of the compact object such as orbital parameters, masses and, when possible, to probe the existence of an event horizon or star's surface. Unambiguous observational probes that serve to this purpose are: {\itshape i).-} Thermonuclear bursts and pulsations (only observed in NSs, \citealp{2006csxs.book..113S,2008ApJS..179..360G} and references therein), and {\itshape ii).-} Mass estimates based on radial velocity measurements (for NS M$_{\star}$ $\leq$ 3 M$_{\odot}$ and M$_{\star}$ $\geq$ 3 M$_{\odot}$ for BH systems). In the particular case of IGR J17091-3624, no thermonuclear bursts or pulsations are observed in its light curves and no direct measurement of its mass has been reported so far. The fact that its variability in outburst resembles the one observed in the BH GRS 1915+105 has been often accepted as strong evidence to classify it as a BH LMXB. However, the NS source MXB 1730-335 has also shown similar X-ray variability \citep[][]{Bagnoli2015}, probing that this kind of variability is unrelated to the nature of the accretor. The similarities with GRS 1915+105 have been used in previous works to constrain the mass of compact object inside IGR J17091-3624, but while some authors place the mass of the system very close to the NS regime (M $\leq$ 5 M$_{\odot}$, see e.g. \citealp{Altamirano2011} and \citealp{Rao2012}) other studies favor the BH scenario by suggesting a higher mass for this source (between 8.7 - 15.6 M$_{\odot}$, \citealp{Rebusco2012,Altamirano2012,Iyer2015}). 
To further investigate the nature of the accretor in IGR J17091-3624, we considered the possible contribution of a thermal component in the X-ray spectra. Assuming that IGR J17091-3624 hosts a NS, we can analyse its X-ray emission at low luminosities to account for the thermal emission associated to the NS surface. Since we have stated in Section \ref{XMM-Newton_fits} that multi-component models are not well constrainted due to the lowest count rates in our spectra, it is only possible to estimate an upper limit for the neutron star temperature.  We considered canonical values for the NS ($M_{NS}=1.4\:M\odot$, $R_{NS}=10\:km$) and the normalization fixed to unity, with the temperature as the only free parameter during the fitting. For D = 8kpc and D = 35 kpc, we obtained an upper limit for the neutron star temperature (as seen by an observer at infinity) of kT$^{\infty}_{eff}$ $\sim$ 68 eV and kT$^{\infty}_{eff}$ $\sim$ 136 eV,  respectively. It is interesting to note that a value of kT$^{\infty}_{eff}$ $\sim$ 136 eV would be in fact very consistent with the temperature reported for the NS SAX J17508-2900 when this source developed post-outburst 'rebrightenings', similar to those observed in IGR J17091$-$3624, after its 2008 outburst \citep{Parikh2017b}. If this hot thermal component is confirmed, our results would place IGR J17091-3624 among the hottest known NSs and rather peculiar. Although the detection of a thermal contribution in the spectra of LMXBs at their lowest luminosities could be attributed to the NS nature of the accretor, some studies have shown that for several NS systems such as SAX J1808.4$-$3658 and EXO 1745$-$248 the X-ray spectra in quiescence are well described with a pure power-law spectrum. These results would necessarily imply that probing the non-existence of a thermal component might not be enough to exclude a NS system inside IGR J17091-3624 (see e.g. discussion in \citealp{Degenaar2012a}). 

\subsection{The true quiescent state} \label{sec:quiescence}
\begin{figure*}
	\includegraphics[clip,scale=0.30]{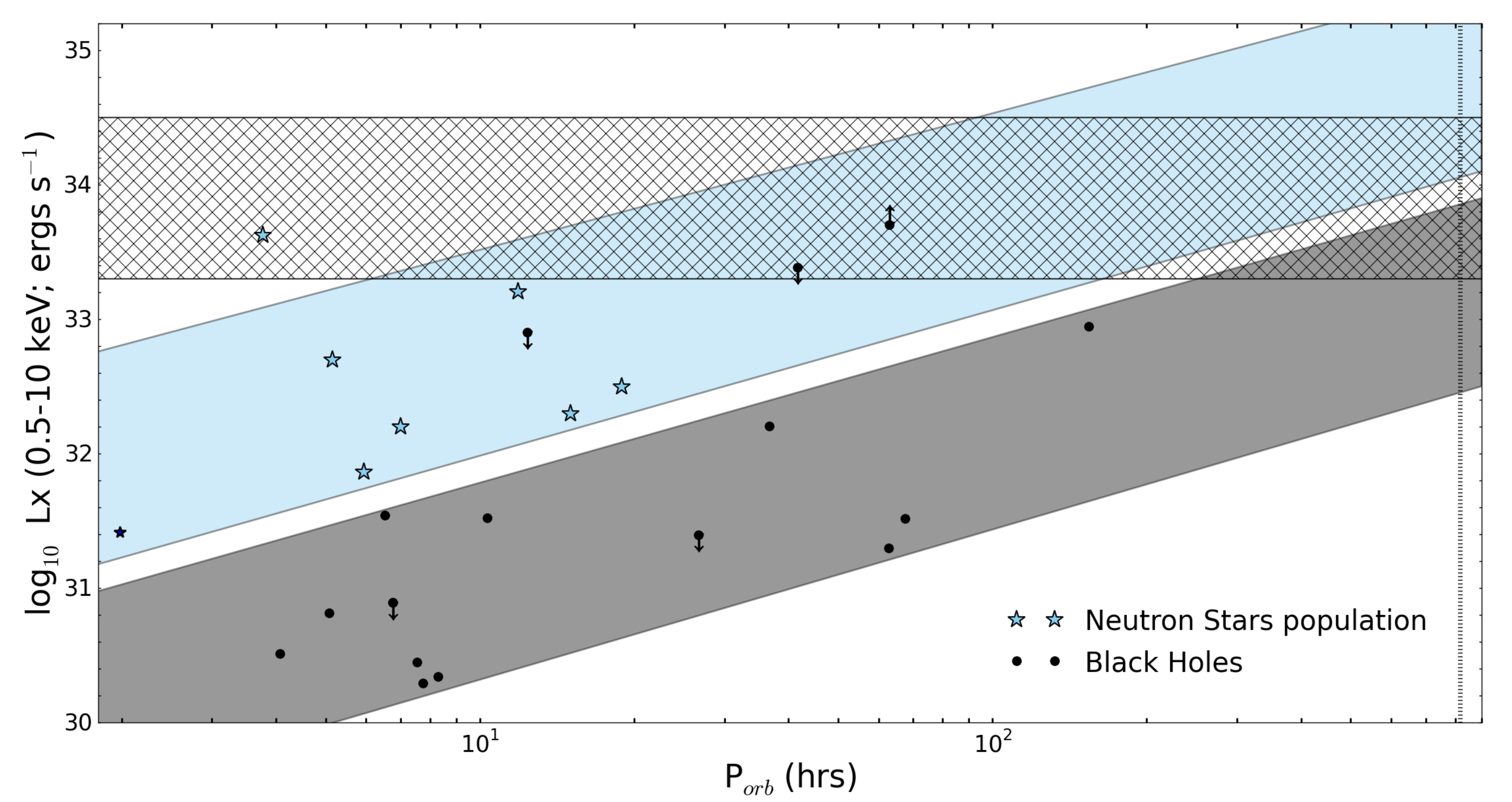}
    \caption{L$_{x}$ vs P$_{orb}$ for confirmed NS (blue stars) and BH (black dots) binaries. Gray and blue shaded areas mark the expected distribution for these two populations within this diagram, according to \citet{Reynolds2011}. We marked with a hatched region the approximate location of IGR J17091-3624. The orbital period of GRS 1915+105 is also indicated, close to the right edge of the plot, with a dotted vertical line.}
    \label{Diss_01}
\end{figure*}

\hspace*{0.5cm} To investigate this we can start by assuming that the lowest fluxes observed do correspond to the quiescent state of IGR J17091-3624, estimate its quiescent X-ray luminosity (L$_{q}$) and compare it with the known distribution of BHs and NSs luminosities in quiescence. Since the distance to IGR J17091-3624 is still unknown, we considered the wide range of distances found in the literature (e.g \citealp{Rodriguez2011,Capitanio2012,Wijnands2012,Janiuk2015}). We estimated L$_{q}$  ranging between 1.7-4.6 $\times$ 10$^{33}$ ergs s$^{-1}$  and 1.1 - 3.4 $\times$ 10$^{34}$ ergs s$^{-1}$ for distances of 8 - 13 kpc and 20 - 35 kpc, respectively. In Figure \ref{Diss_01} we show the X-ray luminosity (L$_{x}$) vs orbital period (P$_{orb}$) of confirmed NS and BH systems at quiescence, taken from \cite{Reynolds2011}. NS and BH systems are plotted with blue and black symbols, respectively. We mark the upper and lower limits with arrows. Based on our estimated L$_{q}$, we indicate the approximate location of IGR J17091-3624 within this diagram. Given that no orbital period has been determined for IGR J17091-3624, we use a horizontal green strip (at the top of the figure) to compare the source with the known NS and BH binaries. If IGR J17091-3624 was in true quiescence at this time, our results would place the source among the few BH LMXBs with significantly high  L$_{q}$. These findings are consistent with those reported by \cite{Wijnands2012}, which pointed out that the high luminosity obtained would entail an orbital period $\geq$ 200 hrs for IGR J17091-3624. They argue that such a long orbital period would favor the BH nature of the binary system, given the orbital period observed in GRS 1915+105. Their results also suggest that the high luminosity found for IGR J17091-3624 could imply that the source has not reached true quiescence at the time of the observations in 2006-2007. Similar results are found when we compare IGR J17091$-$3624 with the NS LMXBs population. The estimated L$_{q}$ would place IGR J17091$-$3624 also among the few NS sources, such as EXO 0748$-$676 (the blue star in the green shaded area), with relatively high luminosity in quiescence. Although, based on the evidence of residual accretion in quiescence provided by \cite{2011A&A...528A.150D} and peculiarities pointed out by \cite{Degenaar2011}, EXO 0748$-$676 seems to be more an exception than the rule. Irrespective of the acretor inside IGR J17091$-$3624, an estimate of the normalized Eddington luminosity for distances D = 8 - 35 kpc yields $l_{x}$ = 10$^{-6}$ to 10$^{-5}$ (as defined by \citealp{Plotkin2013} and with M = 8 - 16 M$_{\odot}$ taken from \citealp{Iyer2015}) which is just above the upper limit value expected for the quiescent state ($l_{x}$ $\sim$ 10$^{-5.5}$). Therefore, it is highly likely that we have not seen the source at its true quiescent state so far.

\subsection{Spectral evolution with X-ray luminosity} \label{sec:Dis_Spectral_evolution}

\begin{figure*}
	\includegraphics[clip,scale=0.34]{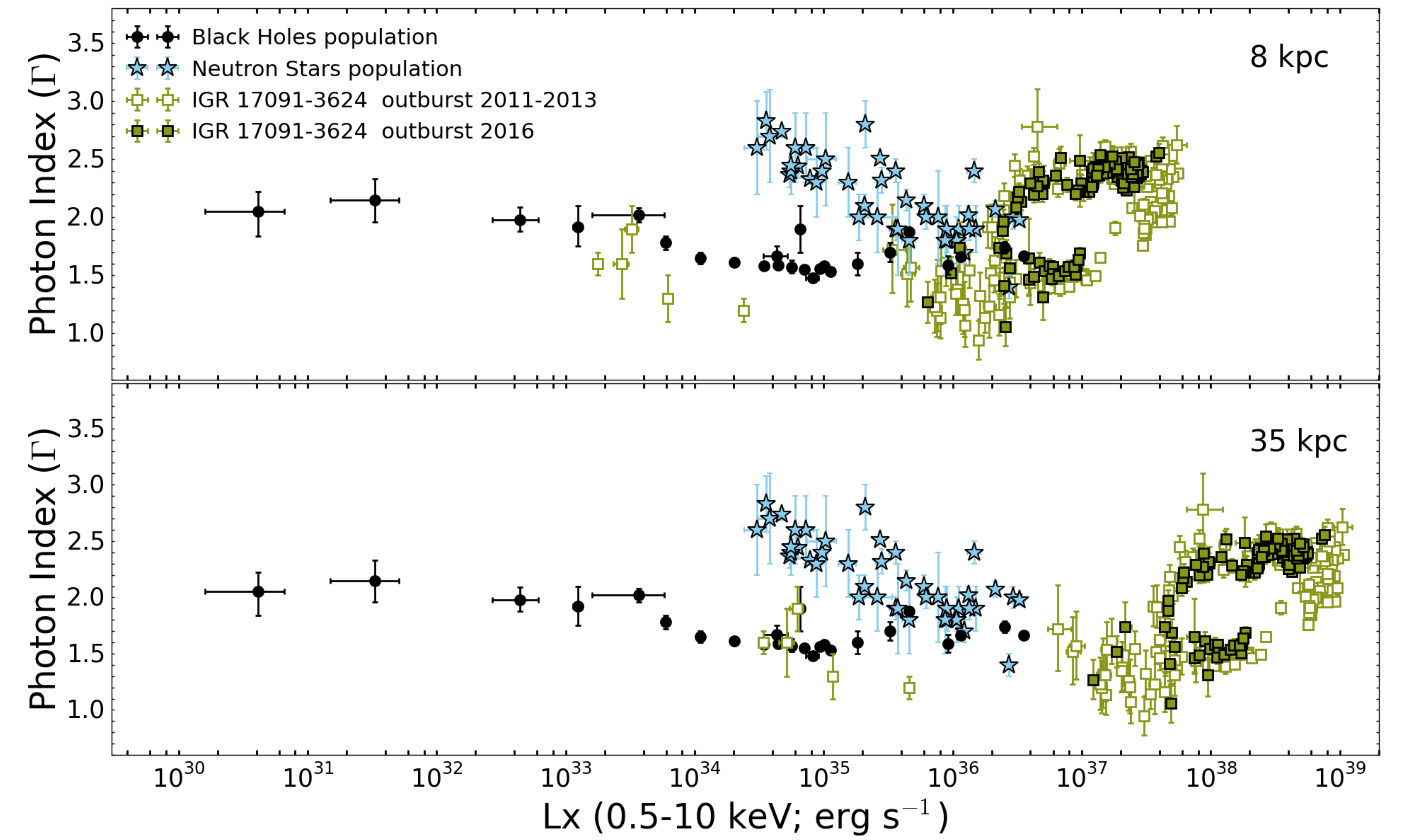}
    \caption{Spectral evolution of LMXBs with decreasing luminosity. Photon index and X-ray luminosities for BHs (black dots) and NS (blue stars) systems taken from \citet{Wijnands2015}. The squares in green indicate the luminosity and photon index values obtained for IGR J17091-3624 at its last two outbursts. We present in the {\itshape top panel} the luminosity values obtained considering $d$ = $8\:kpc$ and at the {\itshape bottom panel}, the values obtained when we assumed $d$ = $35\:kpc$.}
    \label{Diss_02_Histeresis}
\end{figure*}

The analysis of the X-ray emission of different NS and BH binaries demonstrated that, irrespective of the nature of the compact object, their spectra seem to become softer with decreasing luminosities (see e.g. \citealp{Reynolds2014,Plotkin2013}). Moreover, \citet{Wijnands2015} suggested that there is a clear difference in spectral properties of the global population of BH and NS systems at luminosities below 10$^{35}$ ergs s$^{-1}$. In Fig. \ref{Diss_02_Histeresis} we present the photon index ($\Gamma$) and corresponding X-ray luminosity (L$_{x}$) for different LMXBs taken from \cite{Wijnands2015}. A strong correlation between decreasing luminosity and softening of the spectra is observed in both types of LMXBs. But, while the spectra from neutron star systems become softer at relatively high luminosities, $\sim$ 10$^{36}$ ergs s$^{-1}$, for the black holes systems softening is only seen after they reach luminosities below  10$^{34}$ ergs s$^{-1}$. Additionally, the photon index values in the case of BH binaries seem to plateau around $\Gamma$ $\sim$ 2.0 towards quiescence state \citep{Plotkin2013}, whereas higher $\Gamma$ values are found in the case of neutron star systems. To analyse the behaviour of IGR J17091-3624 in the context of the whole population of LMXBs, we have plot in green squares the corresponding L$_x$ and $\Gamma$ obtained from {\itshape Swift}/XRT, {\itshape XMM-Newton} and {\itshape Chandra} data for distances of 8 kpc ({\itshape top panel}) and 35 kpc ({\itshape bottom panel}), respectively. In the high L$_x$ regime, above $\sim$ 10$^{36}$ ergs s$^{-1}$, our results nicely trace an hysteresis loop with the corresponding transitions from hard-to-soft and soft-to-hard states for the outbursts of the source in 2011 (open symbols) and 2016 (filled symbols). It is also seen from this plot that, below $L_{x}$=$\sim$ 10$^{36}$ ergs s$^{-1}$, IGR J17091-3624 luminosity drops by 2 orders of magnitude but its spectra never become softer than 1.7. Under the assumption that softening in the spectra of neutron star binaries is expected to occur at luminosities below 10$^{36}$ ergs s$^{-1}$, we conclude that:
\begin{itemize}
    \item[\textit{i.-}] For IGR J17091-3624 located at a distance of 8 kpc ({\itshape top panel}) the source reached $L_{x}$ as 10$^{33}$ ergs s$^{-1}$ with no evidence of softening in the spectra, which would ruled out a NS nature for the acretor. Nonetheless, the photon index found for IGR J17091-3624 is slightly harder than the expected values for a BH source. 
    \item[\textit{ii.-}] If the source is located at larger distances ({\itshape bottom panel}, at D = 35 kpc), the observed trend seems to be better match the values found for the BH population rather than that of the NS systems. 
\end{itemize}


\section{Concluding remarks}\label{sec:Conclusion}

\hspace{0.5cm} We presented a general description of the X-ray emission of IGR J17091-3624 during the past 10 years as seen by {\itshape Swift}/XRT, {\itshape Chandra} and {\itshape XMM-Newton}. We clearly identified the 2011-2013 outburst showing flare-like behaviour and characterized the low level variability observed from June to October 2013 at a flux level below 5 $\times$ 10$^{-10}$ erg cm$^{-2}$ s$^{-1}$. We observed that the source reached the same flux level at the tail end of the 2016 outburst but showed no variability of any kind. Based on the analysis of the long term {\itshape Swift}/XRT light curve we found that the last two outbursts of IGR J17091-3624 evolved on different timescales.  Using {\itshape XMM-Newton} and {\itshape Chandra} data, we found that the X-ray flux from the source changes by a factor of $\sim$ 5 to 10 at its low luminosity period in 2013, but we did not find strong evidence of spectral softening at this stage which might imply the source has not been observed in its true quiescent state so far. We compared the spectral properties of IGR J17091-3624, during its transition to quiescence, with the spectral behaviour observed in the whole population of LMXBs, and conclude that the acretor inside this object seems to behave more like a BH source rather than a NS.

\section*{Acknowledgements}
We thank the anonymous referee for a thorough reading of the manuscript and helpful comments that led to notable improvements in the quality of this work. MP is a Faculty for the Future Alumnae of Schlumberger Foundation. She appreciates the financial support from Schlumberger Foundation, the COSPAR Capacity Building Workshops Fellowship Programs and CONACyT for current funding in the process of publishing this work. DA acknowledges support from the Royal Society. DA and VAC acknowledges support from the Royal Society International Exchanges ``The first step for High-Energy Astrophysics relations between Argentina and UK". VAC also acknowledges support from the Spanish \textit{Ministerio de Ciencia e Innovaci\'on} under grant AYA2017-83216-P. JC thanks the Science \& Technology Facilities Council and the Royal Astronomical Society for their financial support. ND is supported by a Vidi grant from the Netherlands Organization for Scientific Research (NWO). AP and RW are supported by a NWO Top Grant, Module 1, awarded to RW. This work made use of data supplied by the UK {\itshape Swift} Science Data Centre at the University of Leicester. We thank the {\itshape Chandra} Help Desk team for their guidance and suggestions using the data reduction and analysis software. 

\section*{DATA AVAILABILITY}

The datasets underlying this article are available in the domain:https://heasarc.gsfc.nasa.gov







\appendix

\bsp	
\label{lastpage}
\end{document}